\documentclass[aps,pra,twocolumn,floatfix,10pt,nofootinbib]{revtex4}
\usepackage{epsfig,psfrag,nicefrac}
\usepackage{mathrsfs}
\usepackage{amsmath}

\newcommand{\ie}{i.e.}
\newcommand{\eg}{e.g.}
\newcommand{\ii}{\mathrm{i}}
\begin{document}
\title[Preparation and measurement of coherence]{Optical preparation and measurement of atomic coherence at gigahertz bandwidth}
\author{Paul Siddons, Charles S Adams and Ifan G Hughes}
\affiliation{Department of Physics, Durham University, South Road, Durham, DH1~3LE, UK}
\date{\today}

\begin{abstract}
\noindent We detail a method for the preparation of atomic coherence in a high density atomic medium, utilising a coherent preparation scheme of gigahertz bandwidth pulses.  A numerical simulation of the preparation scheme is developed, and its efficiency in preparing coherent states is found to be close to unity at the entrance to the medium.   The coherence is then measured non-invasively with a probe field.  
\end{abstract}
\maketitle

\noindent 

Controlling the absorptive and dispersive properties of high density alkali metal vapours has allowed the realisation of storage of light~\cite{Phillips2001}, quantum memory~\cite{Julsgaard2004} and entanglement of macroscopic systems~\cite{Jensen2011}.  The preparation of atomic coherence continues to draw a lot of attention~\cite{Karpati2004,Demeter2010,Pustelny2011}, with potential applications in quantum computation and quantum information protocols~\cite{Bouwmeester2000}.   The control of quantum states required in quantum logic operations is often achieved via the application of optical fields to atomic systems, for example, in the implementation of qubit rotations for logic gates~\cite{Kis2002}. 
In this instance the qubit takes the form of an atomic coherence state, but usage of single photons as `flying qubits' is also an area of great interest since transmitting information via light is common place in telecommunications~\cite{Lucamarini2011}.  Often both atomic and photonic qubits are utilised and hence the requirement of reliable transfer of quantum states between photons and atoms, which is the principle behind optical quantum memory~\cite{Lvovsky2009,Moiseev2011}.  The phenomenon of photon echoes~\cite{Kurnit1964,Rubtsova2007} has been used to store and retrieve arbitrary single-photon wave packets~\cite{Moiseev2001} along with the related gradient echo memory~\cite{Hosseini2011,Carreno2011}.
The storage and retrieval of photons at  high speeds (gigahertz bandwidth) for optical quantum memory~\cite{Reim2010,Reim2011} is advantageous for quantum computation, which calls for high-rate operations such as fast quantum gates based on Rydberg atoms~\cite{Urban2009,Gaetan2009}.

Most quantum computation schemes require a few working states amongst which interactions are mediated via optical control fields, and typically require the system to be prepared initially in a pure state.
Preparation schemes based on spontaneous emission, such as Coherent population trapping (CPT)\cite{Zhu1999}, take many excited state lifetimes for a medium to reach its final state, which substantially limits the bandwidth of the operation.  Also, they begin to fail at increasingly high density, as photons scatter from the pumping field and continue to interact with the sample~\cite{Fleischhauer1999}. 
For these reasons, transfer via an off-resonant coherent mechanism is preferred, in which there is a certain amount of freedom to choose the final atomic state of the atom, and can be completed over timescales faster than those necessary for incoherent pumping.  Stimulated Raman adiabatic passage (STIRAP) is one such technique for the highly efficient transfer of population between two non-degenerate metastable states, facilitated via a stimulated two-photon transition involving an unstable intermediate state~\cite{Bergmann1998}.  The technique has been further expanded with the generalisation of the three levels into degenerate manifolds~\cite{Kis2004}.  In this case adiabatic transfer for any arbitrary pure or mixed initial ground state can be achieved under certain conditions.

Probe pulses at the single-photon level have been demonstrated in atomic vapour~\cite{Hockel2010}: to operate at low-light levels strong atom-light coupling is required.  The interaction can be enhanced by using an ensemble containing many atoms, and often necessitates the use of a high density medium~\cite{Kubasik2009}.  This brings with it some technical issues as the preparation of atomic states in such media can be challenging.  Of particular note is the large attenuation and distortion suffered by resonantly interacting light as it propagates through the medium.  Preparation schemes normally rely on specific field amplitude and phase, and if these conditions are not maintained throughout the medium, an inhomogeneous sample will be produced.  STIRAP has also been investigated in situations were back action on the light from the medium is taken into consideration, \eg\ for optically thick media~\cite{Kozlov2009}.  Here evolution of the fields as they propagate through the medium can result in the break down of the conditions required for STIRAP, resulting in incomplete or non-adiabatic population transfer.

In this paper we are not interested in complete transfer out of the ground-state manifold, but rather selective removal of certain states of an incoherent mixture which will lead to an increase in the coherence of the ground-state.  Our aim is to produce a partial coherence in an optically thick medium, and then proceed to measure the degree of coherence by observing the polarisation rotation of a weakly interacting pulse.  The use of off-resonant dispersive measurements is less invasive than resonant, absorptive processes, and leads to the possibility of quantum nondestructive (QND)~\cite{Grangier1998} and `weak'~\cite{Aharonov1988,Hosten2008} measurements.
The structure of the remainder of this paper is as follows: in section~\ref{sec:atomlight} we describe the atom-light system necessary for the preparation scheme and provide details of the numerical model used in the simulation; in section~\ref{sec:prep} we show the result of the simulation; section~\ref{sec:model} details a model of polarisation rotation of a weak probe, the results of which are given in section~\ref{sec:rot}.  Finally we demonstrate the optimal conditions for the STIRAP process in section~\ref{sec:STIRAP}, before drawing our conclusions in section~\ref{sec:conc}.

\section{Atom-light system}\label{sec:atomlight}

\begin{figure}[htb]
\centering
\includegraphics*[width=8.5cm]{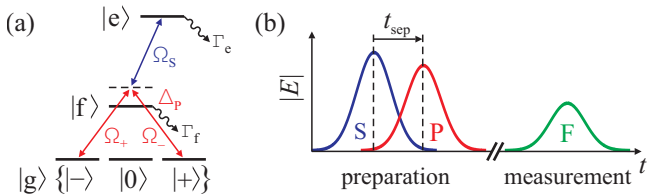}
\caption[Schematic of the stimulated Raman adiabatic passage scheme]{(a)  Energy level scheme.  (b) Pulse sequence: the mutual interaction of the Stokes (S) and pump (P) pulses with the medium constitute the preparation stage. The two pulses are separated in time by $t_{\mathrm{sep}}$, which may be negative.  The preparation stage is succeeded by the measurement stage, achieved via a relatively weak Faraday pulse.}
\label{fig:levels}
\end{figure}

We consider the energy level scheme shown in figure~\ref{fig:levels}(a).  This consists of $|\mathrm{g}\rangle$, a $J_{\mathrm{g}}=1$ ground state which is a manifold of degenerate magnetic sublevels $\{|-\rangle,|0\rangle,|+\rangle\}$, an intermediate state $|\mathrm{f}\rangle$ with $J_{\mathrm{f}}=0$, and an excited state $|\mathrm{e}\rangle$.  This three-level system is referred to as a ladder or cascade system, in which the levels successively increase in energy. For simplicity we do not consider the degeneracy of the excited state.  The excited state population decays at a rate $\Gamma_{\mathrm{e}}$, a fraction of which reaches the intermediate state.  Similarly, the intermediate state population decays at a rate $\Gamma_{\mathrm{f}}$, where it is distributed equally among the ground state sublevels.  As is typical for the STIRAP process, the excited state population is relatively long-lived.  The ground-intermediate state coupling  is via the pump electric field $\mathbf{E}_{\mathrm{P}}$, with associated slowly-varying\footnote{Slowly-varying quantities are denoted by a tilde ($\sim$) throughout this paper.} envelope $\tilde{\mathbf{E}}_{\mathrm{P}}=\tilde{E}_+\boldsymbol{\epsilon}_++\tilde{E}_-\boldsymbol{\epsilon}_-$. Here we have written the polarisation state in the helical basis, where the components $\tilde{E}_+$ and $\tilde{E}_-$ stimulate the $|-\rangle\leftrightarrow|\mathrm{f}\rangle$ and $|+\rangle\leftrightarrow|\mathrm{f}\rangle$ transitions, respectively.   The Faraday field $\mathbf{E}_{\mathrm{F}}$ stimulates the same transitions as the pump, but is of much lower intensity.  Note that the remaining ground-state sublevel $|0\rangle$ is only coupled to the other states via incoherent decay processes.     Intermediate-excited state coupling is via the Stokes field $\mathbf{E}_{\mathrm{S}}$.  The strength of a particular pump/Faraday-mediated transition is defined via the Rabi angular-frequency  $\Omega_\pm=c_{m_J}\mathbf{d}_{\mathrm{gf}}\cdot\tilde{\mathbf{E}}_\pm/\hbar$. Here $\mathbf{d}_{\mathrm{gf}}$ is a reduced dipole matrix element describing the dipole coupling strength of a particular $|L\rangle\rightarrow |L'\rangle$ transition; the coefficients $c_{m_J}$ are factors governing the strength of specific $|J,m_J\rangle\rightarrow|J',m'_J\rangle$ transitions.  Likewise, the Stokes-mediated transition has a Rabi angular-frequency $\Omega_{\mathrm{S}}=c_{m_J}\mathbf{d}_{\mathrm{fe}}\cdot\tilde{\mathbf{E}}_{\mathrm{S}}/\hbar$.  The pump (Faraday) fields are detuned from resonance by $\Delta_{\mathrm{P(F)}}=\omega_{\mathrm{P(F)}}-\omega_{\mathrm{fg}}$; the Stokes field is detuned from resonance by $\Delta_{\mathrm{S}}=\omega_{\mathrm{S}}-\omega_{\mathrm{ef}}$.  The pulse sequence for the experiment is shown in figure~\ref{fig:levels}(b).

Having described how energy levels are linked via the applied fields, we obtain a resultant atomic Hamiltonian, under the rotating-wave approximation, of
\begin{eqnarray}
\mathscr{H}_{\mathrm{RWA}}&=&-\frac{1}{2}\hbar\Bigl(\Omega_+|\mathrm{f}\rangle\langle-|+\Omega_-|\mathrm{f}\rangle\langle+|+\Omega_{\mathrm{S}}|\mathrm{e}\rangle\langle\mathrm{f}|\nonumber\\&&+2\Delta_{\mathrm{P}}|\mathrm{f}\rangle\langle\mathrm{f}|+2(\Delta_{\mathrm{S}}+\Delta_{\mathrm{P}})|\mathrm{e}\rangle\langle\mathrm{e}|\Bigr)+\mathrm{H.c.},
\end{eqnarray} 
where $\mathrm{H.c.}$ denotes the Hermitian conjugate.
In order to simulate the atom-light interaction, the electromagnetic fields are modeled as classical plane waves which co-propagate along the quantisation axis $z$.  The medium is modeled as an ensemble via the density operator $\hat{\rho}$~\cite{CohenBook1998}, whose diagonal matrix elements $\rho_{mm}$ give the probability of an atom occupying state $|\mathrm{m}\rangle$, while the off-diagonals $\tilde{\rho}_{mn}$ give the level of coherence between states $|\mathrm{m}\rangle$ and $|\mathrm{n}\rangle$. The master equation for the time dependence of $\hat{\rho}$ is given by
\begin{eqnarray}
\frac{\partial\hat{\rho}}{\partial t}&=&\frac{\ii}{\hbar}[\hat{\rho},\mathscr{H}_{\mathrm{RWA}}]\nonumber\\&&+\Gamma_{\mathrm{f}}\sum_{m=+,0,-}\hat{\sigma}_{m\mathrm{f}}\hat{\rho}\hat{\sigma}_{m\mathrm{f}}^{\dagger}-\frac{1}{2}\{\hat{\rho},\hat{\sigma}_{m\mathrm{f}}^{\dagger}\hat{\sigma}_{m\mathrm{f}}\}\nonumber\\&&+\Gamma_{\mathrm{e}}\left(\hat{\sigma}_{\mathrm{fe}}\hat{\rho}\hat{\sigma}_{\mathrm{fe}}^{\dagger}-\frac{1}{2}\{\hat{\rho},\hat{\sigma}_{\mathrm{fe}}^{\dagger}\hat{\sigma}_{\mathrm{fe}}\}\right),
\label{eq:master}
\end{eqnarray}
where the curly brackets
$\{\}$ denote the anticommutator, and $\hat{\sigma}_{mn}=c_{mn}|m\rangle\langle n|$ is the lowering operator.
This master equation needs to be solved simultaneously with the slowly-varying envelope form of the Maxwell wave equation
\begin{eqnarray}
\Bigr(\frac{\partial}{\partial t}+c\frac{\partial}{\partial z}\Bigr)\tilde{E}_{\pm}&=&\ii\omega_{\mathrm{P(F)}}\frac{\mathcal{N}_{\mathrm{a}}c_{m_J}d_{\mathrm{gf}}}{\epsilon_0}\tilde{\rho}_{\mathrm{f}\mp},\label{eq:MB1}\\
\Bigr(\frac{\partial}{\partial t}+c\frac{\partial}{\partial z}\Bigr)\tilde{E}_{\mathrm{S}}&=&\ii\omega_{\mathrm{S}}\frac{\mathcal{N}_{\mathrm{a}}c_{m_J}d_{\mathrm{fe}}}{\epsilon_0}\tilde{\rho}_{\mathrm{ef}}\label{eq:MB2},
\end{eqnarray}  
where~(\ref{eq:MB1}) and (\ref{eq:MB2}) apply to the pump~(Faraday) and Stokes pulse, respectively.  The set of coupled equations~(\ref{eq:master})-(\ref{eq:MB2}) form the well-known Maxwell-Bloch equations (see for example~\cite{AllenEberly1975,Berman2010}), and are solved numerically using a Chebyshev pseudospectral time-domain method~\cite{Cheb,SiddonsThesis}. 

To give an explicit example, we choose  to model the $5\mathrm{S}_{1/2}(F=1)\rightarrow5\mathrm{P}_{3/2}(F=0)\rightarrow5\mathrm{D}_{5/2}$ transition found in $^{87}$Rb.  This system has an intermediate-state decay rate of $\Gamma_{\mathrm{f}}=2\pi(6.065\mathrm{MHz})$, all of the atoms decaying out of state $|\mathrm{f}\rangle$ ends up in the ground state.  The excited state decays at the rate $\Gamma_{\mathrm{e}}=2\pi(0.66\mathrm{MHz})$; only a fraction ($0.65$) of the population decaying from $|\mathrm{e}\rangle$ ends up in  $|\mathrm{f}\rangle$, the remaining fraction decays to other states not included in our five-level system.  The reduced dipole matrix element of the pump transition $d_{\mathrm{gf}}=5.177ea_0$~\cite{Siddons2008}, with transition coefficients $c_{m_J}=\nicefrac{1}{3}$ for the three transitions  $\{|-\rangle,|0\rangle,|+\rangle\}\leftrightarrow|\mathrm{f}\rangle$; for the Stokes transition
the reduced dipole element and transition coefficient are $d_{\mathrm{fe}}=1.262ea_0$~\cite{Heavens1961} and $c_{m_J}=-\sqrt{\nicefrac{3}{10}}$  (here $e$ is the magnitude of the charge of an electron, $a_0$ is the Bohr radius).  The Gaussian pulses have a full-width at half-maximum (FWHM) $\delta t=1$~ns, with a $15\pi$ area for both components of the pump pulse and the Stokes pulse.  The pump is initially linearly polarised at $-\pi/4$~rad to the $x$-axis, and has a detuning $\Delta_{\mathrm{P}}=2\pi(10~\mathrm{GHz})$ from resonance.   We assume two-photon resonance between the Stokes and pump fields, requiring that $\Delta_{\mathrm{P}}+\Delta_{\mathrm{S}}=0$.   The Faraday pulse has an area of $10^{-3}\pi$, is initially right-circularly polarised and is 5~GHz detuned from resonance. The medium has an atomic density of $\mathcal{N}_{\mathrm{a}}=10^{20}$~m$^{-3}$ (corresponding to a vapor temperature of $\sim150^\circ$C).

\section{Preparation of the medium}\label{sec:prep}

Before simulating the preparation stage of the atom-light interaction, we first examine the effect of quantum interference~\cite{Ficek2005} in the system.  The two competing paths to the excited state $|-\rangle\rightarrow|\mathrm{f}\rangle\rightarrow|\mathrm{e}\rangle$ and $|+\rangle\rightarrow|\mathrm{f}\rangle\rightarrow|\mathrm{e}\rangle$
lead to quantum interference: a well-known phenomena in the interaction of multi-state systems with coherent light, see for example the review article~\cite{Fleischhauer2005}.  It is instructive to transform the bare-atom set of basis states in to a new set which takes into account the interaction with the light fields.  We reformulate the ground state manifold $\{|-\rangle,|0\rangle,|+\rangle\}$ using the Morris-Shore (MS) transformation~\cite{MorrisShore1983} to $\{|\mathrm{b}\rangle,|0\rangle,|\mathrm{d}\rangle\}$.  The new basis states (the so-called dressed-atom states~\cite{CohenBook1998}) \begin{eqnarray}
|\mathrm{b}\rangle&=&\frac{1}{\Omega_{\mathrm{P}}^*}(\Omega_+^*|-\rangle+\Omega_-^*|+\rangle),\label{eq:b}\\ |\mathrm{d}\rangle&=&\frac{1}{\Omega_{\mathrm{P}}}(\Omega_-|-\rangle-\Omega_+|+\rangle),\label{eq:d}
\end{eqnarray}
are, respectively, coupled and uncoupled from the state $|\mathrm{f}\rangle$; the magnetic sublevel $|0\rangle$ remains uncoupled.  For simplicity we take the polarisation state of the pump to be fixed throughout the experiment, and thus the dressed states also remain fixed.  The Hamiltonian of the transformed system
\begin{eqnarray}
\mathscr{H}_{\mathrm{RWA}}'&=&-\frac{1}{2}\hbar\Bigl(\Omega_{\mathrm{P}}|\mathrm{f}\rangle\langle\mathrm{b}|+\Omega_{\mathrm{S}}|\mathrm{e}\rangle\langle\mathrm{f}|\nonumber\\&&+2\Delta_{\mathrm{P}}|\mathrm{f}\rangle\langle\mathrm{f}|+2(\Delta_{\mathrm{S}}+\Delta_{\mathrm{P}})|\mathrm{e}\rangle\langle\mathrm{e}|\Bigr)+\mathrm{H.c.},
\end{eqnarray}
shows that the transition $|\mathrm{b}\rangle\leftrightarrow|\mathrm{f}\rangle$ is mediated by the Rabi angular-frequency $|\Omega_{\mathrm{P}}|=\sqrt{|\Omega_+|^2+|\Omega_-|^2}$.  This tells us that the atomic state $|\mathrm{b}\rangle$ is associated with the polarisation state of the pump field $\mathbf{E}_{\mathrm{P}}$;  similarly, the uncoupled state $|\mathrm{d}\rangle$ is associated with a field orthogonal to $\mathbf{E}_{\mathrm{P}}$, the magnitude of which is zero in the preparation stage.  We discuss the implications of this later.   

If we begin with an atomic ensemble in a mixed state, the initial ground state density operator in the MS transformed basis is  $\hat{\rho}_{\mathrm{initial}}=\frac{1}{3}(|\mathrm{d}\rangle\langle \mathrm{d}|+|0\rangle\langle0|+|\mathrm{b}\rangle\langle \mathrm{b}|)$, \ie\ the three possible states are evenly populated and there is no coherence amongst them.  The effect of the preparation fields is to affect a two-photon transition between the states $|\mathrm{b}\rangle$ and $|\mathrm{e}\rangle$.  The final ground state density operator is then $\hat{\rho}_{\mathrm{final}}=\frac{1}{3}(|\mathrm{d}\rangle\langle \mathrm{d}|+|0\rangle\langle0|+\delta|\mathrm{b}\rangle\langle \mathrm{b}|)$, where $\delta\rightarrow0$ for complete population transfer.  Examining the form of the dressed-atom states in equations~\ref{eq:b} and \ref{eq:d} we see that they are orthogonal and are coherent superpositions of the  bare-atom states $|-\rangle$ and $|+\rangle$.  Therefore asymmetry in the populations of the dressed states leads to an increase in the coherence of the ground-state subsystem $\{|-\rangle,|+\rangle\}$.  The aim of the preparation process is to create a partially coherent ground state and the efficiency of the preparation stage can be parameterised as the degree of coherence $p_{+-}=|\rho_{+-}|/\sqrt{\rho_{++}\rho_{--}}$,  which takes a value from zero (an incoherent mixture) to the maximum allowed value of unity (a pure state).

\begin{figure}[htb]
\centering
\includegraphics*[width=8.5cm]{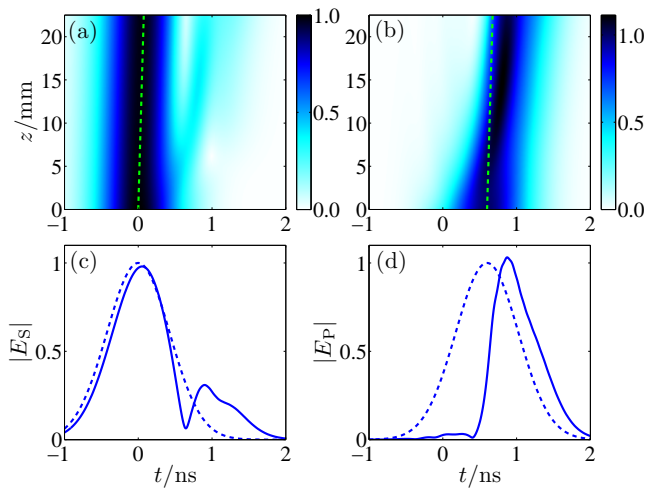}
\caption[Electric field envelopes of the preparation fields]{Electric field envelopes of the preparation fields.  The magnitude of (a) the Stokes and (b) the pump field versus time and displacement inside the medium.  The dashed green line shows the hypothetical position of the peak if the pulse were propagating at the speed of light \textit{in vacuo}.  (c) The Stokes field magnitude versus $t$, at $z=0$ (dashed curve) and $z=22.5$~mm (solid curve); (d) shows the same information for the pump field.}
\label{fig:prep2D}
\end{figure}

\begin{figure}[htb]
\centering
\includegraphics*[width=8.5cm]{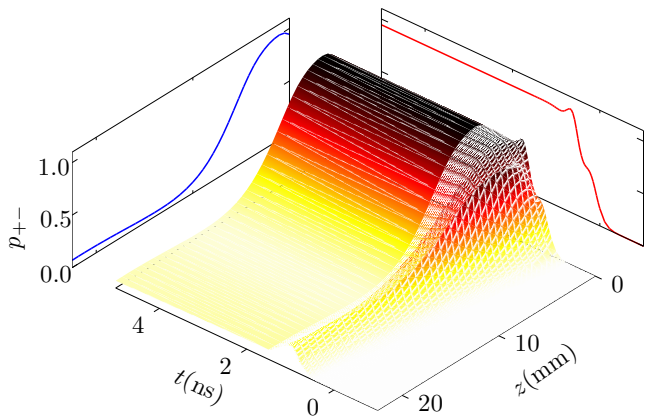}
\caption[The degree of coherence during the preparation stage]{The degree of coherence $p_{+-}$ between the ground state sublevels $|-\rangle$ and $|+\rangle$, shown versus time and displacement inside the medium.}
\label{fig:rho3D}
\end{figure}

Figure~\ref{fig:prep2D} shows the results of pulse propagation in the preparation stage.  In parts (a)-(d) the field envelopes are plotted.  The peak of the Stokes pulse enters the medium at $t=0$~ns, followed by the pump at $t=0.6$~ns.  This pulse spacing amounts to a time separation of one $1/\mathrm{e}$ width, which is the optimal separation for STIRAP using Gaussian envelope functions~\cite{Bergmann1998}.   The front of the Stokes pulse is seen to traverse the medium at close to the speed of light without distortion, due to there initially being no population on the Stokes transition.  As the pump arrives, the two-photon transition can now be affected, leading to strong coupling of the Stokes and pump fields to each other and to the atoms.  Back action on the light distorts the coupled fields as they travel deeper inside the medium.  The initial field parameters were chosen based on the na\"{\i}ve assumption that the relationship between the envelopes of the two preparation pulses remains (relatively) stable.  Unless the atom-light interaction is balanced to maintain the required conditions, the consequence of heavy field distortion is that the STIRAP process is likely disrupted, which is indeed what is seen in figure~\ref{fig:rho3D}.  Here we show the coherence of the subsystem $\{|+\rangle,|-\rangle\}$.  In the first 2~mm the STIRAP process is carried out  with high preparation efficiency, and after the beams have gone ($t>1.5$~ns) the coherent state is of high purity ($p_{+-}>$0.95).  Deeper into the medium, the efficiency is steadily reduced (ignoring the transient `ridge' which is influenced by pulse distortion).  Despite the limited range, STIRAP fares better than preparation via resonant processes (such as inversion via $\pi$-pulses).  Simulations show that during resonant processes the preparation fields are absorbed in the first few tens of microns inside the medium.  

Preparation of a ground-state coherence in optically thick media has been studied previously in reference~\cite{Kozlov2009}.  There, a non-degenerate lambda system was used, \ie\ the final state $|\mathrm{e}\rangle$ is lower in energy than state $|\mathrm{f}\rangle$, in which case the Stokes transition is formally a gain resonance.  Thus in an optically thick medium where back action on the light is significant, energy is transfered from the pump into the Stokes field; this is in contrast to the results of the cascade system seen in this paper where both pump and Stokes fields are absorbed.


\section{A simple model of polarisation rotation}\label{sec:model}

After the STIRAP preparation stage we are left with a medium possessing a ground-state subsystem prepared to a high degree of coherence which is stable against decay (we ignore particle-particle interactions such as collisions~\cite{Weller} which may serve to dephase the partially coherent state).  The rest of the atomic population is found in the metastable excited state from where it will eventually filter back to the ground state, but only over a timescale longer than the duration of the experiment.  To a fair degree of accuracy we can then treat the medium as if it were in a stable state.  In the measurement stage, we apply a weak probing field to the medium.  For a weak enough probe beam the medium is unaffected by the passage of the field, which leaves us free to assume steady-state conditions.

Working in the frequency domain we have a polarisation density of $\tilde{\mathbf{P}}=\frac{1}{2}\epsilon_0\hat{\chi}\tilde{\mathbf{E}}=\mathcal{N}_{\mathrm{a}}\langle\tilde{\mathbf{d}}\rangle$, where $\hat{\chi}$ is the susceptibility tensor.  We can write the field and expectation value of the dipole operator as column vectors, giving the expression
\begin{eqnarray}\hat{\chi}\left(\begin{array}{c}\tilde{E}_+\\\tilde{E}_-\end{array}\right)&=&2\frac{\mathcal{N}_{\mathrm{a}}d_{0}}{\epsilon_0}\left(\begin{array}{c}\tilde{\rho}_{\mathrm{f-}}^{\mathrm{st}}\\\tilde{\rho}_{\mathrm{f+}}^{\mathrm{st}}\end{array}\right).\label{eq:chiE}\end{eqnarray}
Here the dipole matrix element $d_0$ is equal to the reduced dipole matrix element  $d_{\mathrm{gf}}$ multiplied by $c_{m_{J}}$, the relative coefficient of the $\{|-\rangle,|+\rangle\}\rightarrow|\mathrm{f}\rangle$ transitions. Note that the magnitude of $c_{m_{J}}$ is equal for both transitions (due to the symmetry of the electronic wavefunction~\cite{Edmonds1960}).
The steady-state values of the coherence terms can be derived from the Bloch equations, and are found to be
\begin{eqnarray}
\tilde{\rho}_{\mathrm{f-}}^{\mathrm{st}}&=&\frac{\ii d_{0}}{2\hbar}\frac{\tilde{E}_+\rho_{\mathrm{--}}^{\mathrm{st}}+\tilde{E}_-\tilde{\rho}_{\mathrm{+-}}^{\mathrm{st}}}{\Gamma_{\mathrm{f}}/2-\ii\Delta_{\mathrm{F}}}\\
\tilde{\rho}_{\mathrm{f+}}^{\mathrm{st}}&=&\frac{\ii d_{0}}{2\hbar}\frac{\tilde{E}_+\tilde{\rho}_{\mathrm{-+}}^{\mathrm{st}}+\tilde{E}_-\rho_{\mathrm{++}}^{\mathrm{st}}}{\Gamma_{\mathrm{f}}/2-\ii\Delta_{\mathrm{F}}},
\end{eqnarray} 
assuming that the population of the intermediate state $\rho_{\mathrm{ff}}^{\mathrm{st}}\approx0$.  Substituting these steady-state solutions of the coherence into equation~(\ref{eq:chiE}), we can express the susceptibility tensor as 
\begin{eqnarray}
\hat{\chi}=\frac{\mathcal{N}_{\mathrm{a}}d_{\mathrm{0}}^2}{\hbar\epsilon_0}\frac{\ii}{\Gamma_{\mathrm{f}}/2-\ii\Delta_{\mathrm{F}}}\left(\begin{array}{cc}\rho_{\mathrm{--}}^{\mathrm{st}}&\tilde{\rho}_{\mathrm{+-}}^{\mathrm{st}}\\\tilde{\rho}_{\mathrm{-+}}^{\mathrm{st}}&\rho_{\mathrm{++}}^{\mathrm{st}}\end{array}\right).
\end{eqnarray}
In the absence of a coherence between states $|-\rangle$ and $|+\rangle$, the susceptibility tensor is diagonal and thus the field components $\tilde{E}_-$ and $\tilde{E}_+$ propagate independently of one other.  However, in the presence of a coherence this is not the case and there will be interference between the two field components.  The normal modes of the field, \ie\ the field polarisations that propagate independently of each other, can be found by diagonalising $\hat{\chi}$.  However, we noted in Sec.~\ref{sec:prep} that during the STIRAP process the polarisation state of the pump field defines the coupled dressed state $|\mathrm{b}\rangle$ (and the orthogonality condition determines the uncoupled state $|\mathrm{d}\rangle$).  The fields associated with the states $|\mathrm{b}\rangle$ and $|\mathrm{d}\rangle$ are the normal modes of the medium.

A convenient visual representation of light polarisation is the Poincar\'e sphere.  Points in this three-dimensional space correspond to the column vector (here T denotes the transpose operation) $\mathbf{S}=(S_1~S_2~S_3)^{\mathrm{T}}$, the components of which are, respectively, the intensity difference between linearly polarised light in the $x$ and $y$ directions, the intensity difference between linearly polarised light at an angle $+\pi/4$ and $-\pi/4$~rad  to the $x$-axis, and the intensity difference between left and right circularly polarised light.  Note that orthogonality is represented by antipodal points.  A light field with temporally and spatially varying polarisation is generally described by a surface.  The vector/surface is often normalised by the total light intensity, and for fully polarised light each point lies on a sphere of unit radius.  The evolution of the polarisation vector is implicit in the Maxwell-Bloch equations, but to aid the interpretation of the numerical solution to theses equations, we note that the torque equation of motion provides a simple analogy of birefringence~\cite{Huard1997,Genov2011}.  The equation describes the spatial evolution of the polarisation vector $\mathbf{S}$ in response to the anisotropy of the medium, represented by the birefringence vector $\mathbf{a}$:
\begin{eqnarray}
\frac{\mathrm{d}\mathbf{S}}{\mathrm{d}z}&=&\mathbf{a}\times\mathbf{S}.
\label{eq:torque}
\end{eqnarray}
  The geometric interpretation on the Poincar\'e sphere is that $\mathbf{a}$ provides the instantaneous rotation axis and rotary power for the evolution of $\mathbf{S}$.  Note the limitations of this simple picture, however, in that it assumes monochromatic waves in a time-independent medium with zero losses. 
  
The birefringence vector points in the direction of the preponderance of atoms in the $|\mathrm{d}\rangle$ state.  The Stokes parameters of the anisotropy vector can be related to the density matrix elements of the ground-state subsystem via the expression~\cite{Budker2010}
\begin{equation}
\mathbf{a}=\left(-2\mathrm{Re}\left[\tilde{\rho}_{\mathrm{+-}}^{\mathrm{st}}\right]\quad 2\mathrm{Im}\left[\tilde{\rho}_{\mathrm{+-}}^{\mathrm{st}}\right]\quad\rho_{\mathrm{--}}^{\mathrm{st}}-\rho_{\mathrm{++}}^{\mathrm{st}}\right)^{\mathrm{T}}.
\end{equation}
Note the third element in this vector which is due to an imbalance in the populations of the states $|-\rangle$ and $|+\rangle$.  This is the cause of the traditional paramagnetic Faraday effect~\cite{Kubasik2009}, and is a manifestation of circular birefringence.  If the populations are equal and a ground-state coherence exists, the medium will be linearly birefringent \ie\ will respond differently to two orthogonal linearly polarised field components.  In the general case the medium is elliptically birefringent. 

We compare the results of the numerical simulation and torque equation in the next section. 

\section{Results of polarisation rotation}\label{sec:rot}

\begin{figure}[htb]
\centering
\psfrag{a}{(a)}
\psfrag{b}{(b)}
\psfrag{c}{(c)}
\psfrag{d}{(d)}
\psfrag{e}{(e)}
\psfrag{f}{(f)}
\psfrag{g}{$|E_{\mathrm{F}}|$}
\includegraphics*[width=8cm]{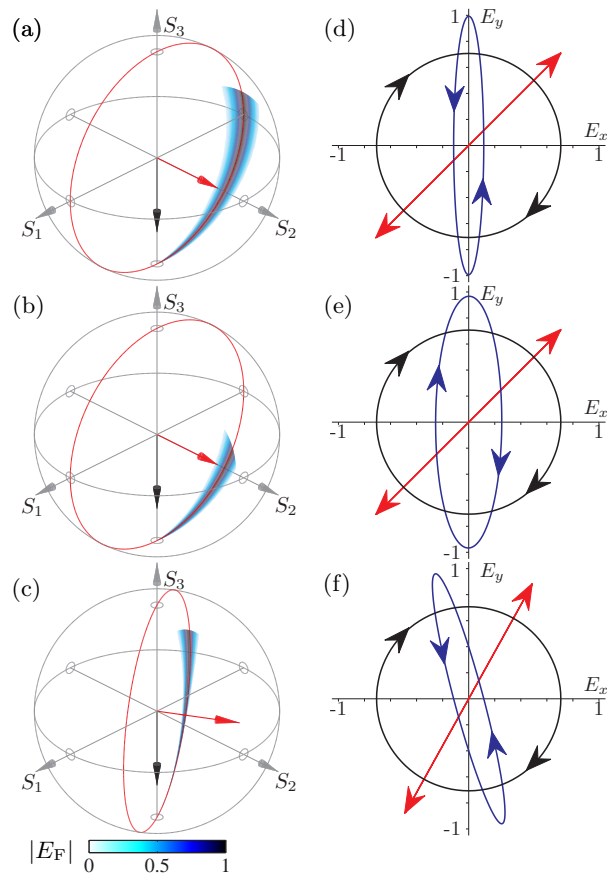}
\caption[Birefringence induced by atomic coherence]{Birefringence induced by atomic coherence. (a)-(c) Poincar\'e sphere representation of the Faraday pulse polarisation state, showing the magnitude of the electric field envelope, $|\tilde{\mathbf{E}}_{\mathrm{F}}|$, as the surface on a unit sphere.  The black arrow represents the initial polarisation state (right-circularly polarised); the red arrow represents the birefringence vector of the medium; the red circle is the path the polarisation state would trace according to the torque model. (d)-(f) polarisation ellipse representation of polarisation, showing the $x$ and $y$ components of $\tilde{\mathbf{E}}_{\mathrm{F}}$ at the entrance (black circle) and exit (blue ellipse) of the medium.  The red line shows one of the normal modes of birefringence.  Arrows represent the circulation of the electric field over one optical period.}
\label{fig:poin}
\end{figure}

In consideration of the complications of the STIRAP process seen in Figs.~\ref{fig:prep2D} and~\ref{fig:rho3D}, a 2~mm long medium will be used to ensure a relatively homogeneous sample for the measurement stage. In this stage, the Faraday pulse interacts with the medium after the preparation fields have exited: this is to avoid further coherent field coupling.  The pulse is sufficiently weak that it does not perturb the medium as it propagates.  As the pulse is detuned far off resonance, it suffers little attenuation/distortion but does experience dispersion, leading to polarisation rotation~\cite{SiddonsNature2009}.  Rotation of the Faraday pulse for different medium parameters is seen in figure~\ref{fig:poin}.  In figure~\ref{fig:poin}(a) the medium is prepared under the same conditions as used in figure~\ref{fig:rho3D}. The Faraday pulse is initially right-circularly polarised, having a polarisation vector $(0~0\,-\!1)^{\mathrm{T}}$. As it propagates through the medium it becomes linearly polarised as it crosses the equator of the Poincar\'e sphere, before becoming left-elliptically polarised.  The polarisation rotates anticlockwise around an axis in the $(0~1~0)^{\mathrm{T}}$ direction, which is to be expected because the medium has population balanced in favor of the state $|\mathrm{d}\rangle$, which is associated with the field polarised at $\pi/4$~rad.  However, the pulse doesn't rotate as a single entity, rather the variation of dispersion over its bandwidth leads to differential rotation. Thus for each position inside the medium the Stokes vector varies in time.  This manifests itself upon the Poincar\'e sphere as the spreading out from a single point. Figure~\ref{fig:poin}(d) shows the field on the polarisation ellipse.  Here the polarisation state corresponding to the peak of the pulse is shown at the entrance and exit of the medium, along with one of the normal modes of the medium (the other mode is orthogonal to this).  

We have previously considered the regime in which the ground state is stable against dephasing mechanisms which have the tendency to lessen the degree of coherence of the atomic subsystem we are interested in.  To model the effects of dephasing we now add a 27~MHz dephasing term to the master equation (this rate is chosen so as to significantly affect the coherence during the few nanoseconds that is the duration of the simulation).  Dephasing causes the sublevel coherence to decay towards an incoherent mixture and thus the rotary power of the birefringent medium decreases with time, though the axis of rotation remains pointing in the same direction of the Poincar\'e sphere.  The Faraday pulse thus rotates to a lesser degree than the case where dephasing mechanisms are ignored, as seen in figure~\ref{fig:poin} parts (b) and (e).

Finally, we consider the effect of an energy difference between the two sublevels involved in the coherence.  This simulates an applied magnetic field used in Faraday rotation.  The energy difference causes precession of the  atomic spin, analogous to the Larmor precession of magnetic moments around an applied magnetic field~\cite{Budker2004}.  The degeneracy of the levels is broken to such an extent that the partially coherent state precesses at a rate of 27~MHz.  By the time the Faraday pulse enters the medium, the birefringence vector has rotated to a new direction, as observed in figure~\ref{fig:poin}(c) and (f).

\section{Effect of varying the pump-Stokes pulse separation time}\label{sec:STIRAP}
\begin{figure}[tb]
\centering
\includegraphics*[width=8.5cm]{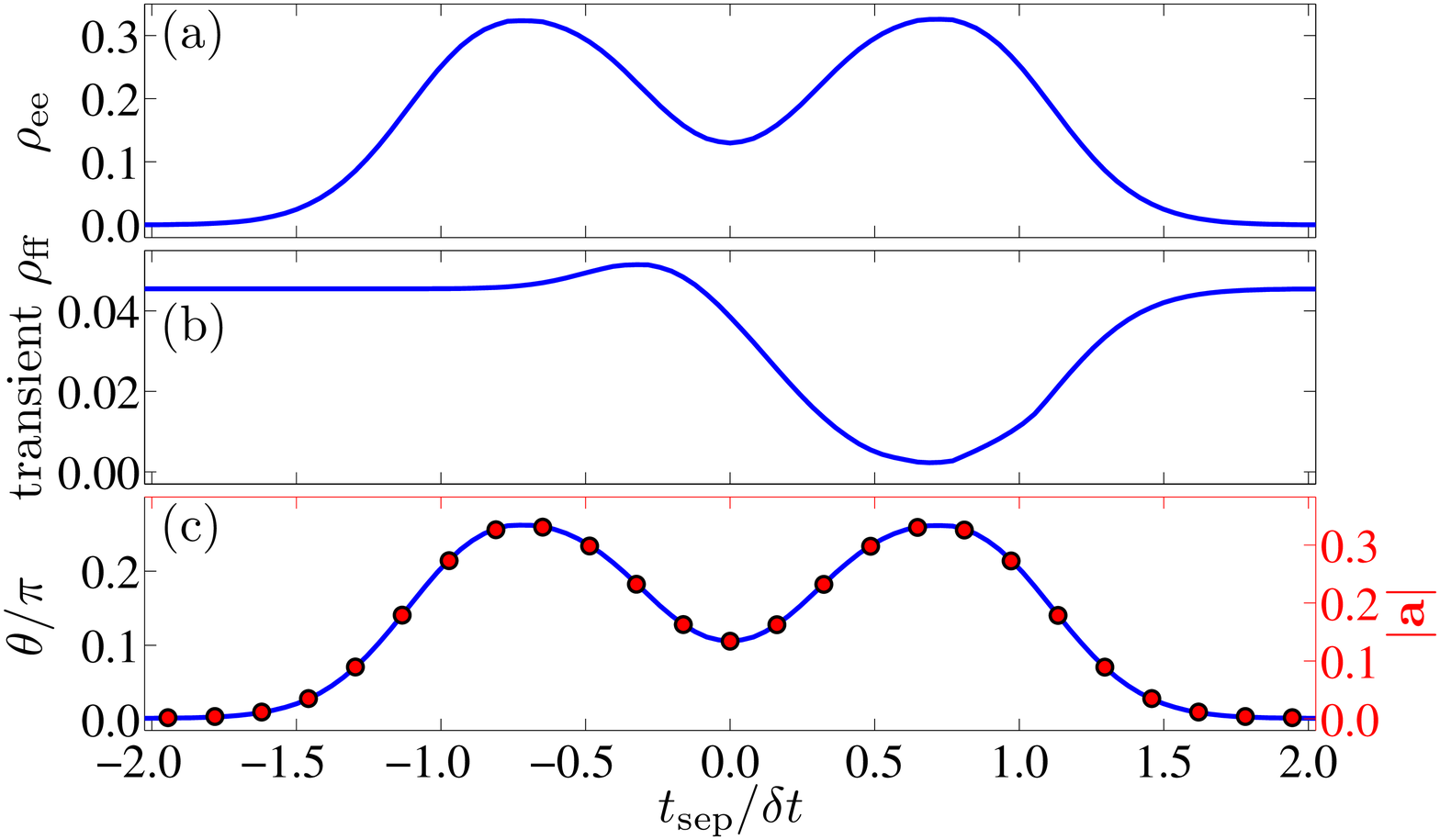}
\caption[Effect of time-separation of the pump \& Stokes pulses on the STIRAP process]{Effect of time-separation of the pump \& Stokes pulses on the STIRAP process. (a) Population of the excited state upon completion of the preparation stage, shown against the separation time between the Stokes and pump pulses.  (b) The maximum population seen in the intermediate state.  The asymmetry around $t_{\mathrm{sep}}=0$ and the trough on the positive side is the signature of the STIRAP process.  (c)  Rotation angle $\theta$ of the Faraday pulse is shown (left axis, solid line), overlaid with the magnitude of the birefringence vector (right axis, data points, the number of which has been reduced for clarity).}
\label{fig:tsep}
\end{figure}
It is well known that the overlap of the Stokes and pump envelopes plays a part in the efficiency of the STIRAP process~\cite{Bergmann1998,Praveen2011}.  Figure~\ref{fig:tsep}(a) shows the population transfered to the excited state versus the separation time between the peaks of the incident Stokes and pump pulses.  For positive $t_{\mathrm{sep}}$, the Stokes pulse precedes the pump, which is the correct order for adiabatic population transfer.  This can be seen from the trough in figure~\ref{fig:tsep}(b), where transfer to the intermediate state is at a minimum.  Note that due to the pump beam being detuned off resonance, the intermediate state is only transiently populated during the preparation stage.  The asymmetry isn't mirrored in figure~\ref{fig:tsep}(a) because by  carrying out the transfer faster than the decay rate we are less harshly punished for going on an excursion to the intermediate state.  Figure~\ref{fig:tsep}(c) shows both the length of the birefringence vector $\mathbf{a}$ and the rotation $\theta$ around this vector experienced by the peak of the Faraday pulse. The magnitude of the birefringence vector and the rotation angle are clearly linked.

\section{Conclusion}\label{sec:conc}

We have demonstrated a theoretical method for the preparation and measurement of a coherence in the ground state of a high density atomic medium, using gigahertz bandwidth pulses.  With the use of realistic parameters, our method is readily amenable to experimental investigation.


\end{document}